\documentclass[aps,prl,superscriptaddress,showpacs,twocolumn]{revtex4}
\usepackage{amsmath,amssymb}

\usepackage{graphicx}

\newcommand{\be}{\begin{eqnarray}}
\newcommand{\ee}{\end{eqnarray}}

\def\ben{\begin{equation}}
\def\een{\end{equation}}
\def\bena{\begin{eqnarray}}
\def\eena{\end{eqnarray}}

\newcommand{\pp}{\partial}

\begin{document}

\title{New Model of Inflation with Non-minimal Derivative Coupling of Standard      
    Model Higgs Boson to Gravity}

\author{Cristiano Germani}
\email{cristiano.germani@lmu.de}
\affiliation{Arnold Sommerfeld Center, Ludwig-Maximilians-University, Theresienstr. 37, 80333 Muenchen, Germany}

\author{Alex Kehagias}
\email{kehagias@central.ntua.gr}
\affiliation{Physics Division, National Technical University of Athens, 15780 Zografou Campus,  Athens, Greece}

\begin{abstract}
In this letter we show that there is a unique non-minimal
derivative coupling of the Standard Model Higgs boson to gravity
such that: it propagates no more degrees of freedom than General
Relativity sourced by a scalar field, reproduces a successful
inflating background within the Standard Model Higgs parameters
and, finally,
does not suffer from dangerous quantum corrections.

\end{abstract}

\pacs{98.80.Cq, 14.80.Bn, 04.50.Kd}

\maketitle

\section{Introduction}
The latest cosmological data \cite{wmap}  agree impressively well with the assumption that our Universe is, at large scales, homogeneous, isotropic and spatially flat, {\it i.e.}, that it is well described by a Friedmann-Robertson-Walker (FRW) spatially flat geometry. This observation is however a theoretical puzzle. A flat FRW Universe is in fact an extremely fine tuned solution of Einstein equations \cite{dodelson}. In the last twenty years or so many attempts have been put forward to solve this puzzle (see for example \cite{many}).
However, the most developed and yet simple idea still remains Inflation \cite{guth}. Inflation solves homogeneity, isotropy and flatness problems in one go just by postulating a
rapid expansion of the early time Universe post Big Bang.

A phenomenological way to achieve Inflation has been pioneered by considering a ``slow rolling'' scalar field \cite{chaotic} with canonical or even
non-canonical kinetic term \cite{k} and lately by non-minimally coupled p-forms \cite{pnflation,proceeding}.
Nevertheless, a fundamental realization of Inflation is still eluding us.

The most economical and yet fundamental candidate for the Inflaton is the Standard Model Higgs boson.
Unfortunately though, the Standard Model parameters are such that no ``slow rolling``
Inflation is possible with the Higgs boson, if minimally coupled to gravity \cite{chaotic}.
To {\it save} the Higgs boson as an Inflaton candidate, in \cite{sha} has been postulated a non minimal
coupling of the Higgs field to gravity. However, during Inflation, the unitarity bound of the theory seems to be violated by non-renormalizable operators emerging from the non-minimal coupling \cite{dub,hert} (see also \cite{mc} for a debate on these results). If instead gravity is non-minimally coupled to derivatives of the scalar field, as shown in this letter, the unitarity bound is not exceeded during Inflation. Then, since the range of parameters in which inflationary attractors exist is greatly expanded in non-minimally derivative coupled scalar field theories \cite{amendola}, there are hopes to define a framework in which the Higgs boson would act as the primordial Inflaton.

In this letter, we show that the unique non-minimally derivative
coupled Lagrangian of the Higgs boson to gravity, propagating no more degrees of freedom than General Relativity sourced by a scalar field, reproduces a successful inflating background within the Standard Model Higgs parameters without unitarity bound violations.

\section{Higgs Boson as Inflaton: a no-go result}
The tree-level Standard Model Lagrangian for the Higgs boson minimally coupled to gravity is
\be
S=\int d^4x\sqrt{-g}\left[\frac{R}{2\kappa^2}-\frac{1}{2}D_\mu {\cal H}^\dag D^\mu {\cal H}-\frac{\lambda}{4}\left({\cal H}^\dag {\cal H}-v^2\right)^2 \right]\ ,\nonumber
\ee
where $R$ is the Ricci scalar, $\kappa$ the gravitational coupling,
${\cal H}$ the  Higgs boson doublet,
$D_\mu$ the covariant derivative with respect to $SU(2)\times U(1)$
and finally $v$ is the vev of the Higgs in the broken phase of the Standard Model.
In the spirit of chaotic Inflation \cite{chaotic}, we will assume that during Inflation no
 interactions with gauge fields are turned on and that the Higgs field is ``large``
 with respect to its vev. With these assumptions we can work with the simpler action
\be
S=\int d^4x\sqrt{-g}\left[\frac{R}{2\kappa^2}-\frac{1}{2}\partial_\mu
\Phi \partial^\mu\Phi-\frac{\lambda}{4}\Phi^4 \right]\ ,\label{a1}
\ee
where for simplicity we introduced a real scalar field $\Phi$ instead of the complex doublet
${\cal H}$ \footnote{There is a subtlety here. Although the replacement of $\cal H$ with $\Phi$
is straightforward for the background, care has to be taken once quantum perturbations are discussed,
see \cite{hert}.}.

To study a FRW solution of this system, we can directly insert into the action the following metric ansatz
\be
ds^2=-N(t)^2dt^2+a(t)^2\delta_{ij}dx^i dx^j\ .\label{metric}
\ee
The only independent Einstein equation, with FRW symmetries, is then recovered by considering the Hamiltonian constraint obtained by varying the action with respect to the lapse $N$ and then setting it to $1$ by time reparameterization invariance \cite{wald}. The field equation for $\Phi$ corresponds instead to the variation of (\ref{a1}) with respect to $\Phi$.

Plugging (\ref{metric}) into (\ref{a1}) we have the following action per unit three-volume
\be
{\cal S}=\int dt a^3\left[-3\frac{H^2}{\kappa^2 N}+\frac{1}{2}\frac{\dot \Phi^2}{N}-N\frac{\lambda}{4}\Phi^4\right] \ ,\nonumber
\ee
where $H\equiv\frac{\dot a}{a}$ is the Hubble constant and $(\dot {})=\frac{d}{dt}$.

The Hamiltonian constraint and field equation are
\be
H^2=\frac{\kappa^2}{6}\left(\dot\Phi^2+\frac{\lambda}{2}\Phi^4\right)\ ,\ \ddot \Phi+3H\dot\Phi+\lambda\Phi^3=0\ .
\ee
Slow roll means $\dot\Phi^2\ll \frac{\lambda}{2}\Phi^4$ together with
\be
|\ddot\Phi |\ll 3H|\dot\Phi|\ ,\label{sr1}
\ee
so that
\be
H^2\simeq \frac{\kappa^2}{12}\lambda\Phi^4\label{HSR1}
\ee
and
\be
\dot\Phi\simeq-\frac{\lambda}{3H}\Phi^3\ .\label{dP1}
\ee
In order to obtain an exponential (de Sitter) expansion of the Universe we need that
\be
-\frac{\dot H}{H^2}\simeq \frac{8}{\kappa^2\Phi^2}\ll 1\ ,\label{c1}
\ee
where the last equality has been obtained by considering (\ref{HSR1},\ref{dP1}).
The slow roll condition (\ref{c1}), requires the field $\Phi$ to be much larger than the Planck scale. This justify our initial assumption to neglect the vev $v$.

A second condition is a compatibility condition.
By combining (\ref{HSR1},\ref{dP1}) we get the following (slightly weaker)
 necessary condition for $\Phi$ satisfying (\ref{sr1}): $\Phi\gg \frac{2}{\kappa}$.

Up to this point it  seems that the Higgs model for Inflation would perfectly work. However, this is, unfortunately, not the case. The framework of Inflation is in fact semiclassical gravity, {\it i.e.}, we can trust the effective inflationary description if and only if curvatures are much smaller than the Planck scales. If this is not the case quantum gravity corrections would play a major role.

During Inflation the curvature scale is proportional to $H^2$. A sufficient condition
to avoid quantum gravity during slow roll is therefore $R\simeq 12 H^2\ll \frac{1}{2\kappa^2}$.
This implies, by using (\ref{HSR1}), $\Phi^4\ll 1/(2\lambda \kappa^4)$.
Combining the previous result with (\ref{c1}), we get $\lambda\ll 10^{-2}$ \cite{chaotic}.
However, the current experimental bounds coming from direct Higgs boson searches as well as from global fit to
electroweak precision data, favors a value of $\lambda$ in the range $0.11<\lambda\lesssim 0.27$
 \cite{bound}, which is obviously incompatible with slow roll Inflation.

\section{Lowering the energy scale during inflation: earlier attempts}

Consider the following non-minimally coupled action \cite{sha}
\be
\!\!\!\!\!\!S=\int d^4x\sqrt{-g} \left[(1+\kappa^2\xi \Phi^2)\frac{R}{2\kappa^2}-\frac{1}{2}\partial_{\mu}\Phi\partial^{\mu}\Phi-\frac{\lambda}{4}\Phi^4\right]\ ,\label{nonminimal}
\ee
where $\xi$ is a parameter.

By a conformal transformation $\tilde g_{\mu\nu}= \Omega^2 g_{\mu\nu}$, where $\Omega^2=(1+\kappa^2\xi\Phi^2)^{-1}$, we can study the system in the Einstein frame. During slow roll ({\it i.e.} neglecting terms in $\dot\Phi$ and $\dot H$), if $\kappa^2\xi \Phi^2\gg 1$, the rescaled action is approximately
\be
S\simeq\int d^4x\sqrt{-g} \left[\frac{\tilde R}{2\kappa^2}-\frac{1}{2}\partial_\mu\tilde\Phi\partial^\mu\tilde\Phi-\frac{1}{4}\frac{\lambda}{\xi^2\kappa^4}\right]\ .\nonumber
\ee
Here we used the canonically normalized field $\tilde\Phi=\ln \Phi/(\kappa\sqrt{\xi})$
and $\tilde R k^2\simeq \lambda\xi^{-2}\ll 1$.
In this case then, with the phenomenological value for $\xi\sim 10^4$ \cite{sha}, it seems that Inflation might be obtained within the Standard
Model value for $\lambda$ without reaching the quantum gravity regime. However,
although quantum gravity is not reached, the unitarity bound of the theory is violated.
This can easily be seen by  going back to the original Jordan frame (\ref{nonminimal}).
In this case (\ref{HSR1}) is modified as $R\simeq 12 H^2\simeq \lambda\Phi_0^2\xi^{-1}$, where $\Phi_0$ is the scalar field value during inflation.

We may now expand the graviton and the Higgs as $\Phi=\Phi_0+\varphi$ and $g_{\mu\nu}=\gamma_{\mu\nu}+\frac{1}{\sqrt{\xi}\,  \Phi_0}h_{\mu\nu}$,
where $\gamma_{\mu\nu}$ is the background metric during the inflationary phase.
The unusual normalization
of the graviton $h_{\mu\nu}$ is to canonically normalize its kinetic term in (\ref{nonminimal}).
 Indeed, during Inflation, in the regime under consideration ($\kappa^2\xi\Phi^2\gg1$), the non-minimal coupling $\xi \Phi^2 R$ dominates over the
 standard
 Einstein $R/\kappa^2$ term. From this expansion we get the non-renormalizable operator
 $\frac{\sqrt{\xi}}{2\Phi_0}\varphi^2\gamma^{\mu\nu}\partial^2 h_{\mu\nu}$, which sets the
  unitarity violation scale of the theory to be $\Lambda=\frac{\Phi_0}{\sqrt{\xi}}$ (for example, by considering the
  $2\varphi\to 2\varphi$ scattering amplitude \cite{dub,hert}).
Imposing that the inflationary energy scale is much below the energy $\Lambda$,
we get the constraint $\lambda\ll 1$. With the Standard model value of $\lambda$,
the energy scale of the inflationary background is so close to
the unitarity bound to challenge the robustness of the Inflationary background against quantum corrections.

\section{A new proposal for the Higgs Inflation}

The insertion of the non-minimal coupling $\Phi^2  R$ in (\ref{a1}), does not introduce
extra degrees of freedom 
as it does not contain higher than two time derivatives. 
This is however not the only possible non-minimal coupling with this property.
In the following, we will show that there is another, unique,  
  non-minimal derivative coupling  of the scalar field to gravity propagating no more degrees of freedom 
than the theory (\ref{a1}). 

Higher curvatures or curvature derivative
coupling  automatically introduce new degrees of freedom. We will therefore only study the following tree-level action
\be
S=\int d^4 x \sqrt{-g}\left[-\frac{1}{2}g^{\mu\nu}\left(1+\zeta R\right)+\frac{w^2}{2} G^{\mu\nu}\right]\pp_\mu\Phi\pp_\nu\Phi\, . \label{ss}
\ee
In (\ref{ss}), $G^{\mu\nu}=R^{\mu\nu}-\frac{R}{2}g^{\mu\nu}$ is the Einstein tensor, $w,\zeta$ inverse mass scales and the positive sign $+w^2$ avoids ghosts propagations. These scalar field interactions to gravity were already considered in the past for early time cosmology. Linear curvatures interactions where studied in \cite{amendola, al} and non linear in \cite{proceeding}. Moreover, the cosmology of a non-minimally coupled Einstein-Yang-Mills-Higgs theory around the Higgs vev, with a cosmological constant, has been studied in \cite{higgscosmo}.

In the ADM formalism \cite{wald}, we can generically decompose the metric as
\be
ds^2=-N^2 dt^2+h_{ij}(dx^i+N^i dt)(dx^j+N^j dt)\ .\label{adm}
\ee
All geometry is thus described by defining a spatial covariant derivative $D_i$ a three-dimensional
curvature ${}^{(3)}R$ (both constructed on $h^{ij}$) and finally an extrinsic curvature
$K_{ij}=\frac{1}{2N}\left(\dot{h}_{ij}-D_i N_j-D_j N_i\right)$. Time evolution is then only related to
 the extrinsic curvature. In General Relativity minimally coupled to a scalar field, in the gauge in
 which $\Phi$ propagates, $N$ is not propagating. 
 
In (\ref{ss}), there is only one
 term containing higher than two time derivatives, {\it i.e.},
 $S_{hd}\sim\zeta \int d^3 xdt \sqrt{h} \,\,  \dot K^i_i\dot{\phi}^2/N^2$.
Clearly, this term increases the number of degrees of freedom of the theory (\ref{ss}), with respect
to General Relativity, by making $N$ a propagating degree of freedom.
To cancel this term one should then take $\zeta=0$. Note that this conclusion is independent
on the foliation (\ref{adm}) chosen.
As a result, the unique non-minimally derivative coupled Higgs theory to gravity, propagating no more degrees of
freedom than General Relativity minimally coupled to a scalar field is
\be
\!\!\!\!\!\!\!S=\!\!\!\int \!\!d^4x\sqrt{-g} \left[\frac{R}{2\kappa^2}-\frac{1}{2}\left(g^{\mu\nu}-w^2
G^{\mu\nu}\right)\partial_\mu\Phi\partial_\nu\Phi-\frac{\lambda}{4}\Phi^4\right].\ \ \label{theory}
\ee
This is the tree-level theory we will discuss in the following. 
En passant, we note that the non-minimal coupling (\ref{theory})  appears in heterotic String Theory for the 
universal Dilaton \cite{dilaton}.

It is easy to see that the non-minimal coupling in (\ref{theory}) may lower
 the effective self coupling of the Higgs boson.  
In a FRW background we have that   $G^{tt}\sim H^2$.
Suppose $w H\gg 1$, $w\gg \kappa$ and that the quantum gravity bound is not exceeded, {\it i.e.}, $12 H^2\ll 1/(2\kappa^2)$. During slow roll ($H\simeq {\rm const.}$) 
we can roughly approximate the action (\ref{theory}) as (we will be more precise later on)
\be
S\simeq\int d^4x\sqrt{-g} \left[\frac{R}{2\kappa^2}-\frac{1}{2}\partial_\mu\bar\Phi\partial^\mu\bar\Phi-\frac{\bar\lambda}{4}\bar\Phi^4\right]\ ,\nonumber
\ee
where the canonically normalized field $\bar\Phi\sim\frac{1}{w H}\Phi$ has been used. In this case, the effective self coupling 
constant is $\bar\lambda\sim\frac{1}{w^4 H^4}\lambda\ll \lambda$.

We may also check if  unitarity bounds are  violated in this case.
For this specific coupling the Hubble equation (\ref{HSR1}) during Inflation is not modified. However the canonical normalization of the scalar field $\Phi$ is. Like before let us again expand around an inflating background. The Higgs and the graviton expands as $\Phi=\Phi_0+\frac{1}{\sqrt{3}\, w \, H}\varphi$, $g_{\mu\nu}=\gamma_{\mu\nu}+\kappa\, h_{\mu\nu}$. The factor in front of $\varphi$ canonically normalize it in the case we are considering, {\it i.e.}, $w H\gg1$.

The first non-renormalizable operator appearing on the expansion of the action (\ref{theory}) is now
\be
I\simeq\frac{\kappa}{2H^2}\partial^2 h^{\mu\nu}\partial_\mu \varphi\partial_\nu \varphi\ ,
\ee
with the (time dependent) unitarity bound $\Lambda(H)\simeq (2H^2/\kappa)^{1/3}$. By requiring that $R\ll\Lambda(H)^2$, we get $H\ll 1/\kappa$. Our postulated coupling is therefore free of unitarity problems during Inflation (assuming that quantum corrections to our tree-level action are suppressed by the scale $\Lambda(H)$). One can also easily show that the same happens after Inflation, {\it i.e.}, the unitarity bound is never violated.

\subsection{Slow roll Higgs Inflation}

We can now discuss the cosmological solution of the theory (\ref{theory}). We use again the metric ansatz (\ref{metric}) in (\ref{theory}) obtaining the following action per unit three-volume
\be
{\cal S}=\int dt\, a^3\left[-3\frac{H^2}{\kappa^2 N}+\frac{1}{2}\frac{\dot\Phi^2}{N}+\frac{3}{2}\frac{H^2w^2}{N^3}\dot\Phi^2-N\frac{\lambda}{4}\Phi^4\right]\ .\nonumber
\ee
The Hamiltonian constraint and field equation are 
\be
H^2=\frac{\kappa^2}{6}\left[\dot\Phi^2\left(1+9H^2w^2\right)+\frac{\lambda}{2}\Phi^4\right]\ ,\cr
\partial_t\left[a^3\dot\Phi\left(1+3H^2w^2\right)\right]=-a^3\lambda\Phi^3\ .\nonumber
\ee
We will ask the solution to obey the following inequalities
\be
H\gg \frac{1}{3w}\ , ~~~ 9H^2w^2\dot\Phi^2\ll \frac{\lambda}{2}\Phi^4\ , ~~~ -\frac{\dot H}{H^2}\ll1\ ,\label{SR2}
\ee
where the last two are the usual slow roll conditions. Of course (\ref{SR2}) must be cross checked afterwords.

With (\ref{SR2}) we find 
\be
H^2\simeq \frac{\kappa^2}{12}\lambda\Phi^4\ ,\label{Hour}
\ee
and $\ddot\Phi+3H\dot\Phi=-4/(w^2 \kappa^2\Phi)$. By considering the extra slow roll condition
\be
|\ddot\Phi|\ll3H|\dot\Phi|\ ,\label{extra}
\ee
we finally get
\be
\dot\Phi\simeq-\frac{4}{3Hw^2\kappa^2\Phi}\ .\label{dotp}
\ee
The quantum gravity constraint $R\simeq 12H^2\ll 1/(2\kappa^2)$ implies
\be
\Phi^4\ll \Phi_M^4\equiv\frac{1}{2\kappa^4\lambda}\ .\label{max}
\ee
We now need to cross check the various constraints (\ref{SR2},\ref{extra}).
As $\lambda<1$, we have the stronger constraint
$\Phi^6\gg 32/(w^2 \kappa^4\lambda)$. Combining this with (\ref{max}), we have $w/\kappa\gg 10\times \lambda^{1/6}$ or, by considering that $\lambda\geq .11$ \cite{bound}, $w/\kappa\gg 7$ .

The number of e-folds during Inflation are $N=\int_{\Phi_i}^{\Phi_f}H/\dot\Phi d\Phi$, where
$i$ and $f$ stands for the initial and final value of the Inflaton during Inflation. By using (\ref{Hour},\ref{dotp}) and taking $\Phi_f\sim 0$, we obtain:
$w/\kappa\simeq 16 N^{1/2} \lambda^{1/4} r^3$, where $r=\Phi_M/\Phi_i$. Considering the bounds on $\lambda$ \cite{bound} and $N\simeq 60$ we get $w/\kappa\sim 10^2 r^3$. Note that a provoking value $w\sim{\rm TeV}^{-1}$ corresponds to the reasonable ratio $r\sim 10^4$.

\section{Conclusions}

In this paper, we investigated whether the tree-level modification of Einstein gravity via a non-minimal coupling of the Standard Model Higgs field to gravity, can
produce a successful slow rolling inflationary background, without violating unitarity bounds. We found a positive
answer to this question. If the kinetic term of the Higgs is indeed non-minimally coupled to the Einstein tensor,
slow roll Inflation is obtained without exceeding the unitarity bound of the theory. Moreover, we showed that the particular
non-minimal coupling postulated here is unique in the the sense that does not propagate more degrees of freedom than General Relativity minimally coupled to a scalar field. 
 
The early time cosmology ends when, by reheating of the Universe, the Higgs settle to its Standard Model vev $v$.
In this regime, thanks to the derivative coupling, General Relativity is fully recovered.

An important issue concerns the radiative corrections to our model. Taking aside the non-renormalizable character of gravity, there are two types of corrections, those of gravitational in origin and those due to the Standard Model fields. Athough a detailed analysis of this has been postponed for future work, by power counting arguments, gravitational corrections should be controlled by $V(\Phi)/(M^2 \Lambda^2)$, where $V$ is the Higgs potential during inflation. These gravitational corrections are therefore subleading, thanks to the unitarity constraint $H\ll \Lambda$. Standard Model fields would instead introduce logharitmic corrections to the Higgs potential which do not spoil the flatness of the Higgs tree-level potential \cite{will}.

\vskip 0.15cm
\acknowledgements
CG wishes to thank Stefan Hofmann, Alberto Iglesias, Alex Pritzel and Filippo Vernizzi for useful discussions and comments. CG also thanks Gia Dvali for important comments on the non-renormalizable operators. CG is sponsored by the Humboldt Foundation. This work is partially supported by the European Research and Training Network MRTPN-CT-2006 035863-1 and the PEVE-NTUA-2009 program.

\end{document}